\documentclass[fleqn,usenatbib]{mnras}

\usepackage{newtxtext,newtxmath}

\usepackage[T1]{fontenc}
\usepackage{ae,aecompl}

\usepackage{graphicx}	
\usepackage{amsmath}	
\usepackage{amssymb}	

\newcommand{\nustar}{\textit{NuSTAR}}

\newcommand{\swift}{{\it Swift}}

\newcommand{\maxi}{{\it MAXI}}

\newcommand{\src}{GRS~1716$-$249}

\title[\src\ \nustar\ Spectral Analysis]{A \nustar\ View of \src\ in the Hard and Intermediate States}

\author[J. Jiang et al.]{
Jiachen Jiang,$^{1,2,3}$\thanks{E-mail: jj447@cam.ac.uk}
Felix F{\"u}rst,$^{4}$
Dominic J. Walton,$^{3}$
Michael L. Parker,$^{4}$
\newauthor
and Andrew C. Fabian$^{3}$
\\
$^{1}$Department of Astronomy, Tsinghua University, Shuangqing Road, Beijing 100084, China\\
$^{2}$Tsinghua Center for Astrophysics, Tsinghua Univerisity, Shuangqing Road, Beijing 100084, China\\
$^{3}$Institute of Astronomy, University of Cambridge, Madingley Road, Cambridge CB3 0HA, UK\\
$^{4}$European Space Agency (ESA), European Space Astronomy Centre (ESAC), E-28691 Villanueva de la Ca\~nada, Spain\\
}

\date{Accepted XXX. Received YYY; in original form ZZZ}

\pubyear{2019}

\begin{document}
\label{firstpage}
\pagerange{\pageref{firstpage}--\pageref{lastpage}}
\maketitle

\begin{abstract}
We present a detailed analysis of the spectral properties of the black hole transient \src, based on the archival \swift\ and \nustar\ observations taken during the outburst of this source in 2016--2017. The first six \nustar\ observations show that the source is in a canonical hard state, where the spectrum is dominated by a power-law continuum. The seventh \nustar\ observation is taken during the intermediate state where both a disc thermal component and a power-law continuum are shown. All of our observations show a broad emission line feature in the iron band and a Compton hump above 10\,keV. We model the broad band spectra using a high density disc reflection model, where the soft X-ray emission in the hard state is interpreted as part of the disc reflection component. This model enables us to constrain the disc density parameter of \src\ in the range of $10^{19}-10^{20}$\,cm$^{-3}$. We only obtain an upper limit of the inner disc radius using high density disc reflection spectroscopy and the results indicate either a non-truncated disc or a slightly truncated disc with $R_{\rm in}\lesssim20$\,$r_{\rm g}$.
\end{abstract}




\section{Introduction}

X-ray binaries (XRBs) are accretion-powered binary systems that are luminous in X-rays, and typically made up of a compact object accreting from a `normal' stellar companion. Many BH XRBs show transient events that are characterised by outbursts in the X-ray band. Some outbursts last for only few weeks \citep[e.g. MAXI~J1659$-$152,][]{munoz11} while some last for years before they return to the quiescent state \citep[e.g. Swift~J1753.5$-$0127,][]{soleri13}. Two main different spectral states are commonly seen during the outbursts of BH XRBs: 1) the hard state where the spectrum is dominated by a power law-shaped continuum; 2) the soft state where the spectrum is dominated by a disc-blackbody emission \citep[e.g.][]{oda71}. In the hard state, the spectrum often shows broad Fe K emission line and a strong Compton hump above 10~keV in addition to the power-law emission that is associated with the corona \citep[e.g.][]{kitamoto90}. These features are very similar with the ones found in Seyfert galaxies despite different BH mass scales \citep[e.g.][]{walton12} and together are referred to as the disc reflection component. In some cases, BH XRBs show intermediate states between the soft and hard states, where a hard power-law continuum and a disc thermal emission component make approximately equal contribution to the total spectrum \citep[e.g.][]{steiner12, tomsick18}.

Previous reflection-based spectral analysis of BH XRB observations commonly found a super-solar disc iron abundance \citep[e.g.][]{garcia18b}. A possible solution to this problem is a high density disc reflection model \citep[e.g.][]{tomsick18, jiang18, jiang19b}. However, a fixed disc density ($n_{\rm e}$=10$^{15}$\,cm$^{-3}$) was typically assumed in previous models. This assumption is appropriate for the disc of a massive supermassive BH \citep[e.g. BH mass $m_{\rm BH}=M_{\rm BH}/M_{\odot}>10^{8}$ and mass accretion rate $\dot{M}>20\%\dot{M}_{\rm Edd}$,][]{shakura73}. At higher disc densities, the temperature of the disc surfaces increases due to stronger free-free absorption. The soft X-ray band of the disc reflection spectrum turns into a blackbody-shaped emission \citep[][]{ross07,garcia16}. When modelling the broad band spectra with a high density reflection model, the continuum flux level of the disc reflection component increases and the inferred iron abundance parameter drops for the spectral fitting purpose \citep{jiang19c}. The blackbody-shaped emission feature in the high density disc reflection model may be able to explain the weak blackbody components commonly seen in the hard state spectra of BH XRBs \citep{jiang19b} and the blackbody-shaped soft excess emission in Seyfert galaxies \citep[e.g.][]{mallick18,garcia19}.

\src\ is a low-mass XRB first discovered in 1993 \citep{ballet93,harmon93}. It reached a peak flux of 1.4~Crab in the 20--100~keV band within five days. By studying the Na~D absorption lines, \citet{della94} obtained an upper limit of $d\approx2.8$\,kpc for the distance of \src. In our work, we adopt $d=2.4$\,kpc, the same value applied for luminosity calculations in \citet{della94}. \citet{masetti96} estimated a lower limit of $m_{\rm BH}>4.9$ for the mass of the central BH by studying the period of the light modulations during the previous outburst. \src\ was then detected in another outburst in 1995 with a simultaneous radio flare \citep{hijellming96}. 

\src\ was found in an outburst 21 years later by \maxi\ on 18th December, 2016 \citep{masumitsu16, negoro16}. By analysing the spectra from the \swift\ monitoring programme of this outburst, \citet{bassi19} found that the source approached the soft state three times but never reached the canonical soft state. Simultaneous radio observations also show \src\ is located on the radio-quiet branch of the X-ray--radio luminosity plane \citep{bassi19}.

In this paper, we conduct detailed spectral analysis of the seven \nustar\ \citep{harrison13} observations of \src\ triggered during the outburst in 2016--2017. Simultaneous \swift\ snapshot observations in the soft X-ray band are considered as well if available. The first three \nustar\ observations did not have simultaneous \swift\ observations because of the sun violation for \swift. In Section\,\ref{red}, we introduce the data reduction processes; in Section\,\ref{data}, we present the broad band spectral modelling by using high density disc reflection model, study the spectral variability, and conduct multi-epoch spectral analysis to obtain the disc viewing angle and the disc iron abundance; in Section\,\ref{dis}, we discuss our spectral analysis results; in Appendix \ref{appen1}, we study the parameter degeneracy when both the disc thermal and the disc reflection components are considered for the hard state observations of \src.


\section{Data Reduction} \label{red}

\subsection{\nustar\ Data Reduction}

We use the standard pipeline NUPIPELINE V0.4.6 in HEASOFT V6.25 to reduce the \nustar\ data. The calibration file version is v20171002. In addition to the science mode (mode 1) data, we also extracted the products from the spacecraft science (mode 6) event files by following \citet{walton16}. The mode 6 events are considered because the star tracker was blinded by the sun during the early observations of \src\ and was unable to provide an aspect solution. Therefore the source coordinate in the sky was calculated using the spacecraft bus attitude solution for mode 6 observations.  
The inclusion of the mode 6 data maximizes the usable exposure of our observations. See Table\,\ref{tab_obs} for the exposure time in two different modes. The mode 1 source spectra are extracted from a circular region with radii of 120 arcsec, and the background spectra are extracted to be circular regions near the source region on the same chip. The mode 6 source spectra are extracted from a circular region with radii of 110--140 arcsec. We generate spectra using the NUPRODUCTS command. The FPMA and FPMB spectra are grouped to a minimum signal-to-noise (S/N) of 6 and to oversample the energy resolution by a factor of 3. The \nustar\ data are modelled over the full 3--78\,keV band in this work.

\subsection{\swift\ Data Reduction}

We use the standard pipeline XRTPIPELINE V0.13.5 to reduce the \swift\ XRT data \citep{burrows05}. The calibration file version is V20171113. The source spectra are extracted from an annulus region to avoid pile-up effects. The inner radius of the annulus is 5 arcsec and the outer radius of the annulus is 40 arcsec. The XRT spectra are grouped to a minimum S/N of 6 and to oversample the energy resolution by a factor of 3. The \swift\ data are modelled over the 1--8\,keV band in this work. We ignore the spectra below 1~keV due to the known calibration issue with the Window Timing mode.

\begin{table*}
	\centering
	\caption{A list of observations considered in this work. The fourth, fifth and seventh columns show the net exposure of \nustar\ Mode 1, Mode 6  and \swift\ XRT observations respectively.}
	\label{tab_obs}
	\begin{tabular}{cccccccc} 
		\hline
		Obs No. & \nustar\ ObsID & Date & Mode 1 & Mode 6 & \swift\ Obs ID & XRT\\
		        &                &      & ks     & ks     &                & ks \\  
		\hline
		1 & 80201034002 & 2016-12-26 & - & 23 & - & - \\
		2 & 80201034004 & 2016-12-31 & 0.4 & 18 & - & - \\
		3 & 80201034006 & 2017-01-05 & 6 & 47 & - & -\\
		4 & 80201034007 & 2017-01-28 & 7 & 6 & 0034924001 &  1.0\\
        5 & 90202055002 & 2017-04-07 & 18  & 1.8  & 0034924029 & 1.7 \\
        6 & 90202055004 & 2017-04-10 & 16  & 1.8  & 0034924031 & 1.9 \\
        7 & 90301007002 & 2017-07-28 & 89 & 10  & 0088233001 & 3.7\\
		\hline
	\end{tabular}
\end{table*}

\section{Data Analysis} \label{data}

In this section, we first present an overview of the spectral transition through the outburst of \src\ in 2016-2017. Second, we conduct detailed broad band spectral modelling using a high density disc reflection model. In the end, we discuss the changes of the disc-coronal geometry and the disc properties during the outburst by conducting a multi-epoch spectral analysis.  

\subsection{Spectral State Transition} 

\begin{figure}
	\includegraphics[width=\columnwidth]{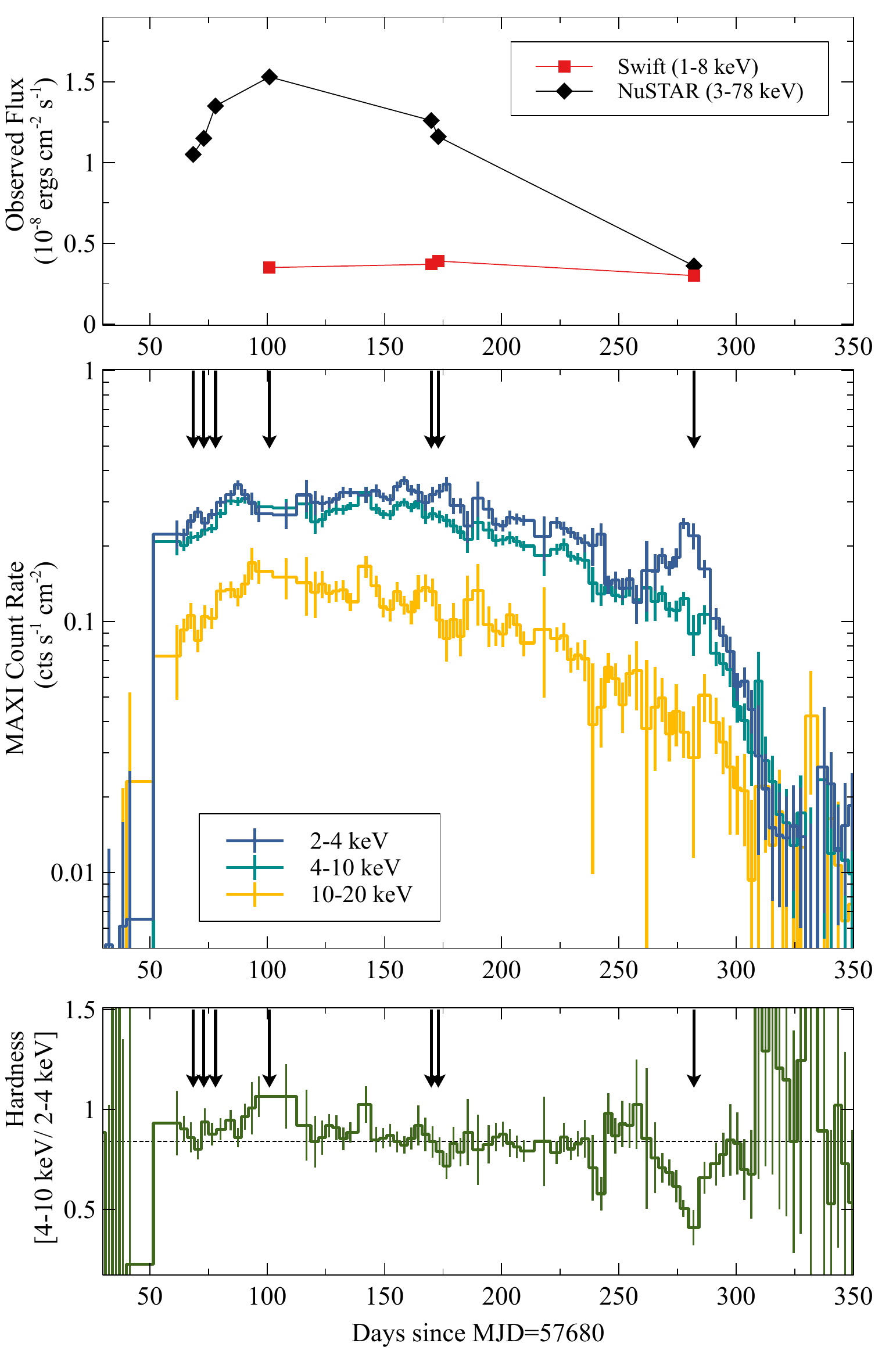}
    \caption{Top: observed X-ray flux in the \nustar\ (black diamonds) and \swift\ (red squares) energy bands. Middle: \maxi\ light curves of \src\ in 3-day bin during the outburst in 2016--2017 (blue: 2--4~keV; red: 4--10~keV; yellow: 10--20~keV). The black arrows mark the dates when the \nustar\ observations were taken. Bottom: the hardness ratio curve obtained by \maxi. The hardness ratio is calculated by the flux ratio between the 4--10~keV and 2--4~keV bands.  }
    \label{fig_maxi_lc}
\end{figure}

\begin{figure*}
    \centering
	\includegraphics[width=17cm]{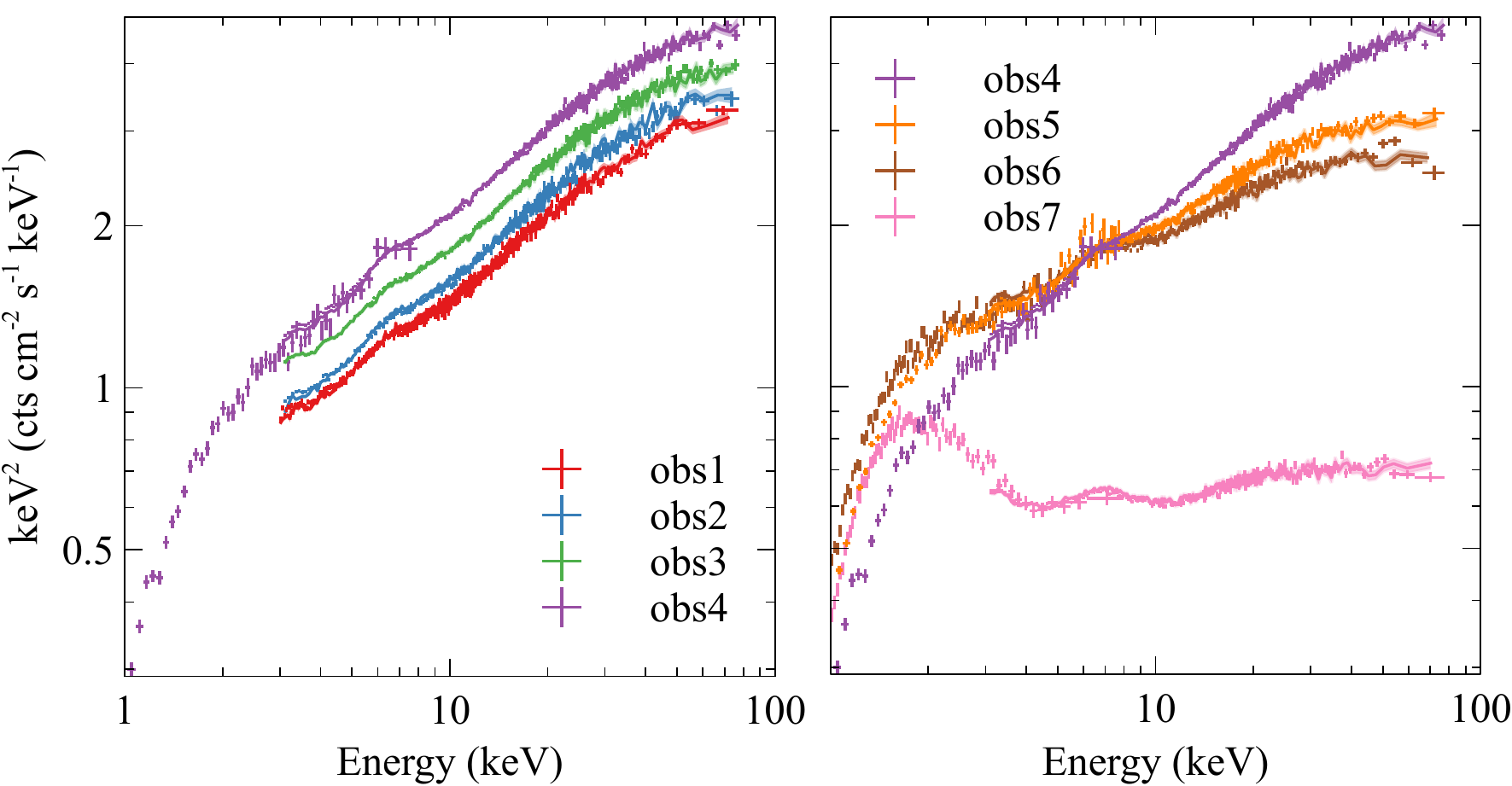}
    \caption{The \nustar\ FPMA (small crosses), FPMB (lines) and \swift\ (big crosses) spectra of obs 1-4 (left) and obs 4-7 (right) unfolded through a constant model. The spectra of obs 1-4 show a similar spectral shape but different flux levels. The \swift\ XRT spectra have been shifted to match the \nustar\ spectra in the overlapping energy band (3--8~keV). The \texttt{constant} model is used to account for the cross-calibration factor during our detailed spectral modelling.}
    \label{fig_eeuf}
\end{figure*}

The observed \nustar\ and \swift\ flux of \src\ is shown in the top panel of Fig.\,\ref{fig_maxi_lc}. The \textit{MAXI} lightcurves of the same period in different energy bands are shown in the middle panel. The outburst of this source lasted for 8 months until it returned to the quiescent state. The X-ray flux rose for the first 2 months of the outburst before starting a slow decay. The bottom panel of Fig.\,\ref{fig_maxi_lc} shows the hardness ratio between the 4--10~keV and 2--4~keV bands. The hardness ratio curve from \maxi\ indicates that the source spent majority of the outburst in the hard state before approaching the soft state near the end of the outburst, although the total X-ray flux continued to decay across this period. The decrease of the hardness ratio is also confirmed by the \nustar\ and \swift\ observations. \src\ returned to the hard state before fading back to quiescence, indicating an outburst with failed transition to the soft state. We refer interested readers to \citet{bassi19} for a complete hardness-intensity diagram (HID) of this outburst obtained by \swift, which monitored \src\ throughout this outburst.

The black arrows in Fig.\,\ref{fig_maxi_lc} mark the dates when the \nustar\ observations were taken during the outburst. The spectra of all the \nustar\ and coordinated \swift\ observations unfolded through a constant model are shown in Fig.\,\ref{fig_eeuf}. The left panel of Fig.\,\ref{fig_eeuf} shows the spectra of the first four observations during the rising phase of the outburst. These four observations show very similar spectral shape but different flux levels. The right panel shows the spectra extracted from the rest of the observations. Obs 5 and 6, which were taken 2 months after the 4th observation, show a softer broad band spectral shape but a similar X-ray flux level compared with the observations in the rising phase of the outburst. Such a change in the spectral shape matches the indication from the \textit{MAXI} lightcurves in Fig.\,\ref{fig_maxi_lc}. Obs 7 was taken 7 months after the beginning of the outburst. The spectra of obs 7 are shown in the right panel of Fig.\,\ref{fig_eeuf} for comparison with previous observations. The last observation shows the lowest total X-ray flux level but a very strong blackbody-shaped soft emission below 3~keV.

\subsection{Spectral Modelling}

\begin{figure}
	\includegraphics[width=\columnwidth]{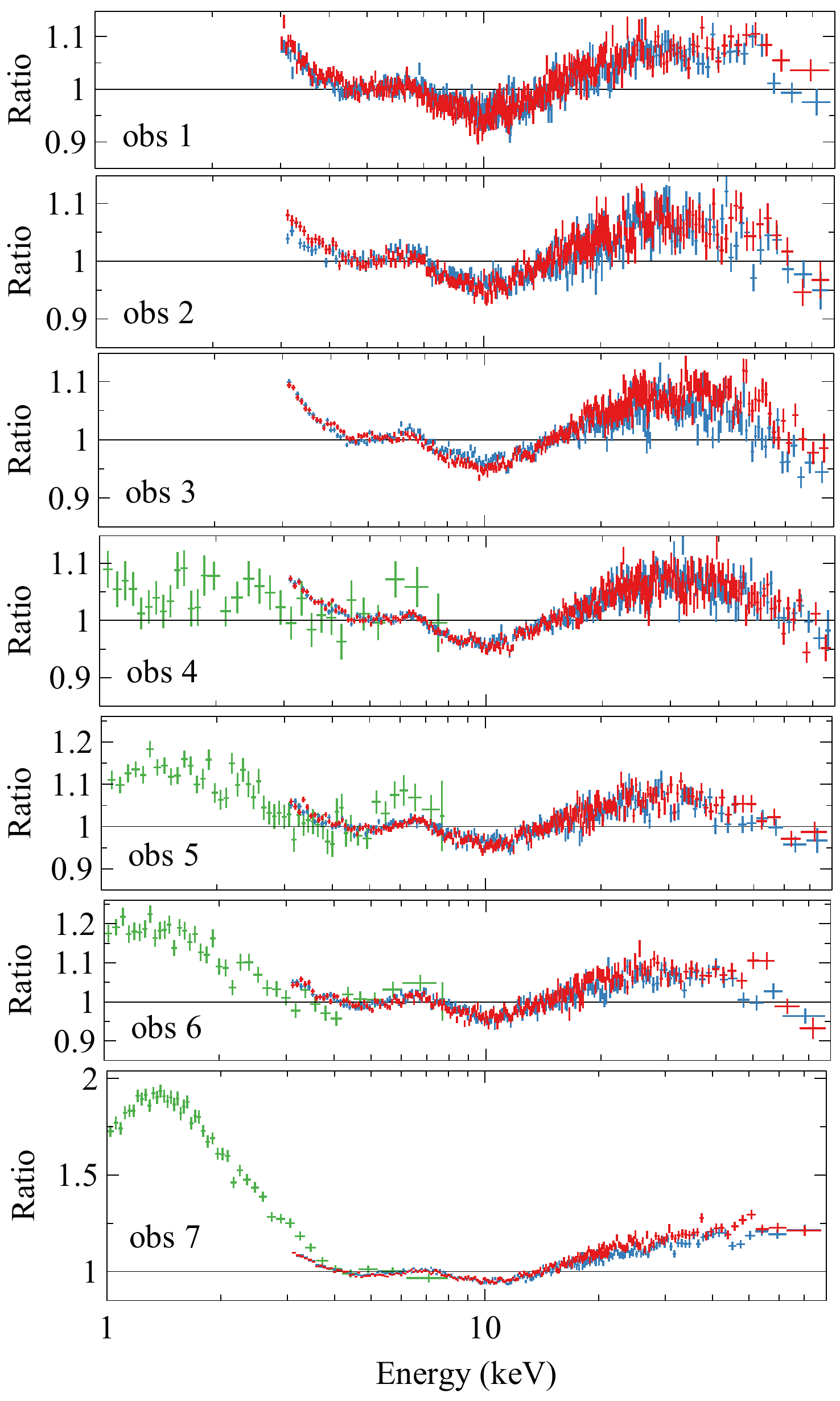}
    \caption{The ratio plots of \nustar\ FPMA, FPMB and \swift\ XRT spectra against the best-fit power-law model in the 3--80~keV band. All the spectra show a broad emission line at 6.4~keV and a strong Compton hump above 10~keV.}
    \label{fig_pl_ratio}
\end{figure}

We use XSPEC V12.10.1b \citep{arnaud96} for spectral analysis and C-stat is considered in this work. The upper limit of the Galactic column density in the line of sight towards \src\ is $3.97\times10^{21}$\,cm$^{-2}$ \citep{willingale13}. We fix the column density of the Galactic absorption at this value when fitting the spectra with absorbed power-law models. Then we model with the column density as a free parameter for the final results.

We first fit the spectra with absorbed power-law models. Data/model ratio plots for these models are shown for each epoch in Fig.\,\ref{fig_pl_ratio}. All the spectra show a broad emission line feature in the iron band and a Compton hump above 10~keV. In addition to these relativstic reflection features, the soft energy band of obs 7 is dominated by a strong excess emission. Based on the evidence for broad Fe K emission and the Compton hump in each of the spectra, we model the broad band spectra with a combination of a coronal emission component and a relativistic disc reflection component. The former component is modelled using the simple power-law model \texttt{cutoffpl} with a high energy exponential cut-off. The latter component is calculated using \texttt{relconvlp * reflionx}. A more developed version of \texttt{reflionx} \citep{ross05, ross07} is used to model the disc reflection model in the rest frame with following parameters: disc iron abundance ($Z_{\rm Fe}$), disc ionization ($\log(\xi)$), disc density ($n_{\rm e}$), high energy cut-off ($E_{\rm cut}$), and photon index ($\Gamma$). The solar abundance in \texttt{reflionx} is calculated by \citet{morrison83}. The ionisation parameter is defined as $\xi=4\pi F/ n$ in units of erg\,cm\,s$^{-1}$, where $n$ is the hydrogen number density and $F$ is the illuminating flux. The high energy cut-off parameter and the photon index in the reflection model are linked to the corresponding parameters of the \texttt{cutoffpl} component. The convolution model \texttt{relconvlp} \citep{dauser13} is used to apply relativistic effects to the rest-frame reflection model. This calculates the emissivity profile according to a simple lamp-post model parameterised by the height of the corona above the disc. Other free parameters in this model are the disc inner radius $r_{\rm in}/r_{\rm ISCO}$ and the disc inclination angle $i$. The spin parameter is fixed at the maximum value 0.998 to fully study the inner disc radius parameter. The outer radius of the disc is assumed to be at 400\,$r_{\rm g}$ as the fits are insensitive to this parameter. The convolution model \texttt{cflux} is used to calculate the flux of each model component between 1--78~keV band. We define the reflection fraction as $f_{\rm refl}=F_{\rm refl}/F_{\rm pl}$, where $F_{\rm refl}$ and $F_{\rm pl}$ are the best-fit flux values given by \texttt{cflux}. Note that this empirical reflection fraction is different from the  physical reflection fraction discussed in \citet{dauser16}. A \texttt{constant} model is used to account for the cross-calibration uncertainity between different instruments. The total model is \texttt{constant * tbabs * (cflux * ( relconvlp * reflionx ) + cflux * cutoffpl ) } (\texttt{MODEL1}) in the XSPEC format.

The combination of a relativistic disc reflection and a power law-shaped coronal emission models offers a very good fit to the spectra of first 6 observations with C-stat/$\nu\approx1.0$, where $\nu$ is the number of degrees of freedom. For example, we obtain a good fit with no evidence for structured residuals and C-stat/$\nu=2665.32/2613$ by modelling the spectra of obs1 with \texttt{MODEL1}. In order to test for any narrow emission line in the iron band, we also tried adding one Gaussian line model \texttt{gauss} with the line energy fixed at 6.4\,keV. This additional line model only improves the fit by $\Delta$C-stat=2 with 2 more free parameters for obs1. Only an upper limit of the equivalent width of the line model is obtained (EW<3\,eV). Similar conclusion can be found when analysing the spectra of obs 2--6. Therefore we conclude that there is no evidence for a distant reflector in our observations.

As shown in Fig.\,\ref{fig_eeuf} and Fig.\,\ref{fig_pl_ratio}, the low energy spectrum of obs 7 shows clear evidence for an additional blackbody-shaped component. By modelling the \swift\ and \nustar\ spectra simultaneously with \texttt{MODEL1}, we fail to obtain a good fit for obs 7. The high density disc reflection model is unable to model the strong `soft excess' emission with C-stat/$\nu$=5250.68/1512. By adding an additional \texttt{diskbb} component, we improve the fit by $\Delta$C-stat=3486 with 2 more parameters. The full model is \texttt{constant * tbabs * (cflux * ( relconvlp * reflionx ) + cflux * cutoffpl + diskbb) } (\texttt{MODEL2}) in the XSPEC format. The requirement for an additional disc-blackbody component agrees with previous analysis \citep[e.g.][]{armas17,bassi19}, marking a transition to an intermediate state where both the disc thermal emission and the coronal emission contribute significantly to the total observed X-ray flux. Similarly, a combination of high density disc reflection and thermal emission from the disc has been also found in the intermediate state of Cyg~X-1 \citep{tomsick18}.

In order to test whether the spectra of the hard state spectra (obs 1-6) require an additional \texttt{diskbb} component, we re-analysed the spectra with \texttt{MODEL2}. \texttt{MODEL2} provides an equivalent fit as \texttt{MODEL1} with C-stat decreasing by 2 and two more free parameters for obs 4 as an example. See Appendix \ref{appen1} for more details about our further Markov chain
Monte Carlo (MCMC) analysis for the hard state observations. We conclude that an additional \texttt{diskbb} component is statistically not required for the spectral fitting for the hard state observations as it does not improve the spectral fitting significantly. However, we do note that there is a degeneracy between the inner disc temperature parameter ($kT_{\rm e}$) of the \texttt{diskbb} model, the disc density parameter in the reflection model, and the column density of the line-of-sight absorption. Similar conclusion has been found in the previous analysis of the hard state observations of GX~339-4 with a variable disc density parameter \citep[][]{jiang19b}.

We also notice that the FPMA and FPMB data of obs 2 and 7 show difference of 6\% in the spectra below 4~keV. Similar disagreement between FPMA and FPMB spectra has also been found in other \nustar\ observations \citep[e.g.][]{madsen17} due to possible calibration uncertainty. 

\subsection{Multi-epoch Spectral Analysis} \label{multi}

\begin{figure*}
	\includegraphics[width=17cm]{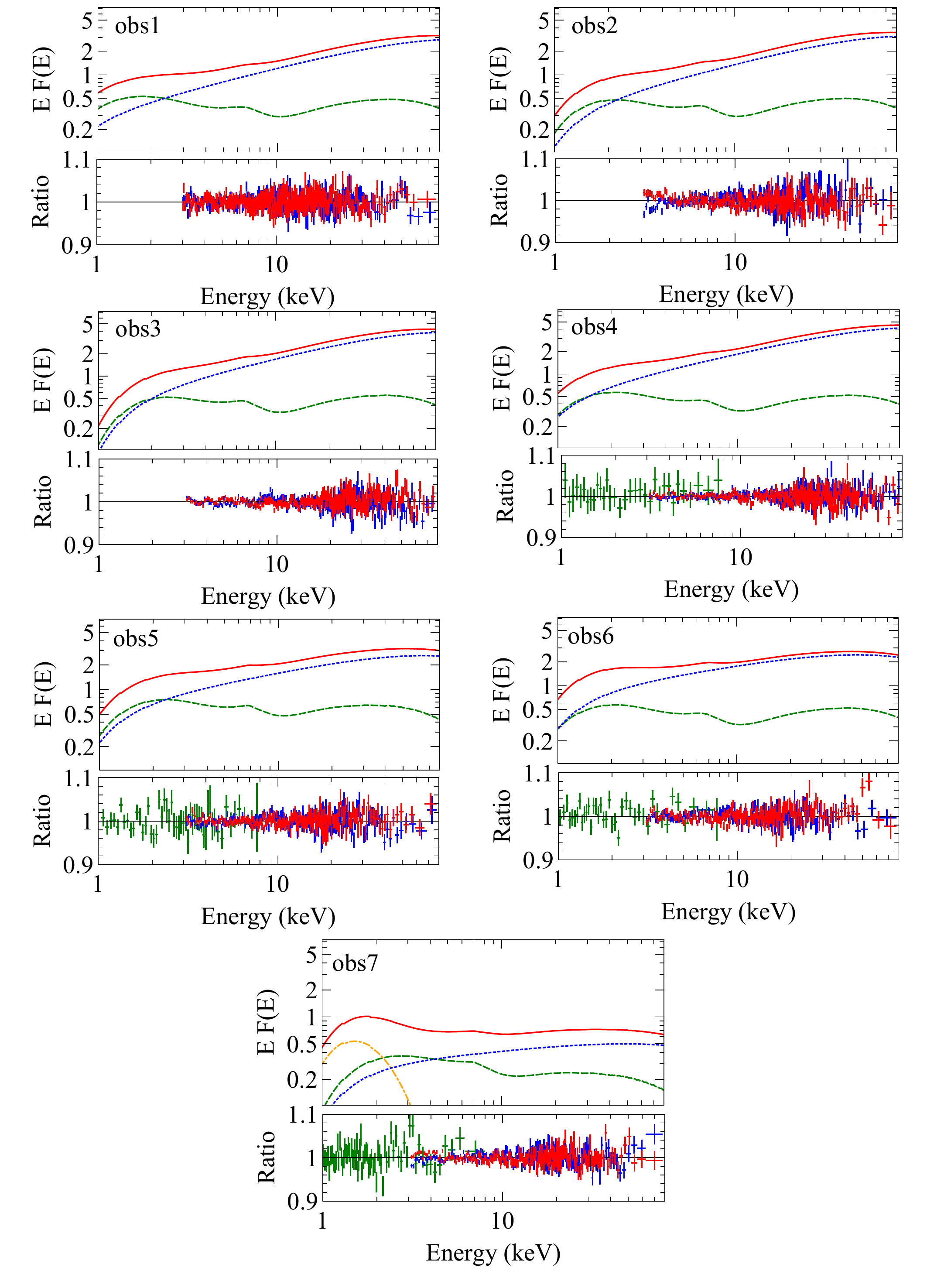}
    \caption{Top panels: the best-fit spectral models (red: total model; blue: coronal emission; green: relativistic reflection; yellow: disc blackbody component). The y-unit is keV$^{2}$(cts\,cm$^{-2}$\,s$^{-1}$\,keV$^{-1}$). Bottom panels: corresponding ratio plots (red: FPMA; blue: FPMB; green: XRT).}
    \label{fig_best_ratio}
\end{figure*}

\begin{figure}
	\includegraphics[width=\columnwidth]{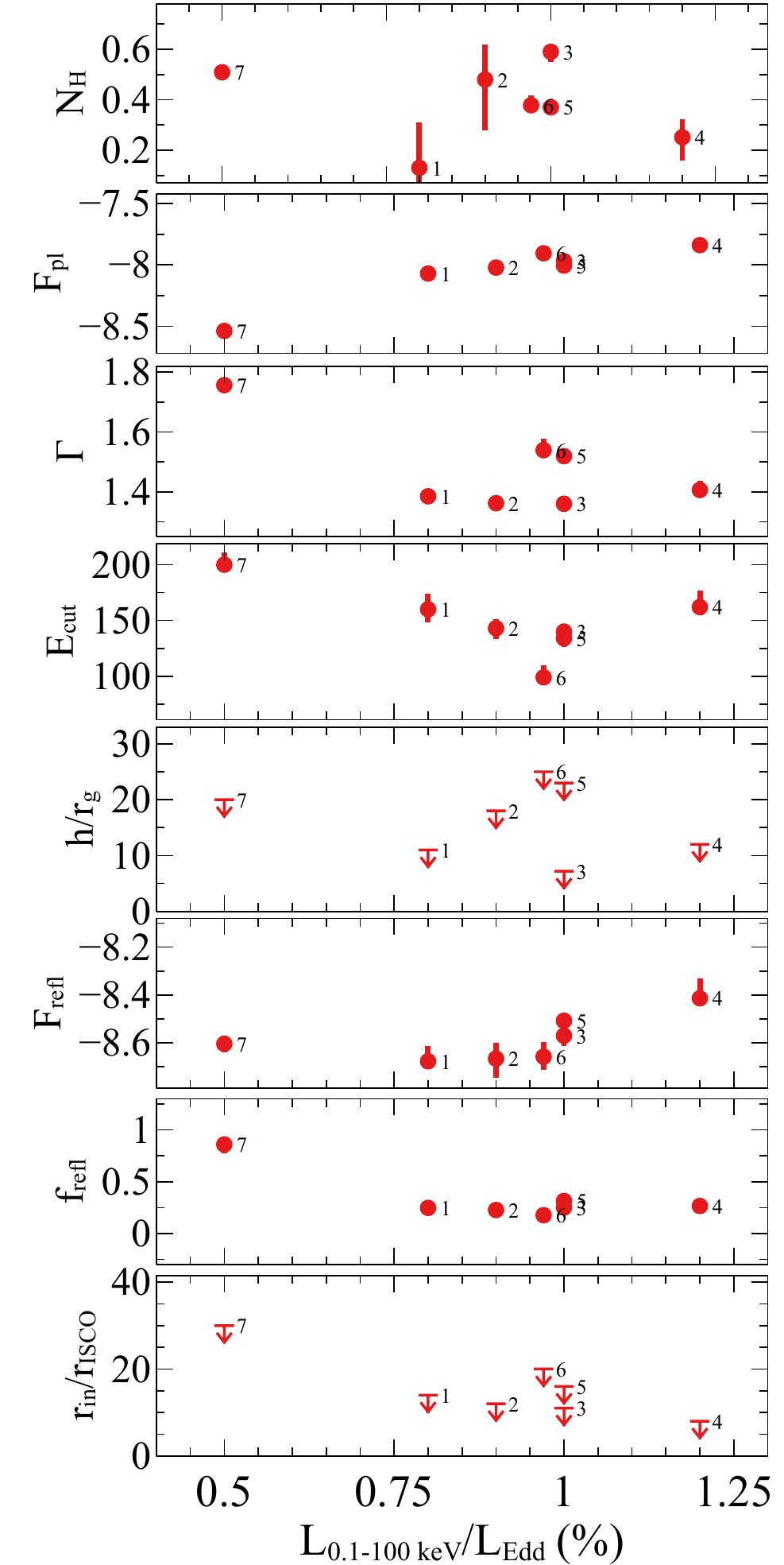}
    \caption{The changes of the best-fit parameter values against the 0.1--100~keV band luminosity in the unit of the Eddington luminosity for a 10$M_{\odot}$ stellar-mass BH. The best-fit parameter values are in the same unit as the ones in Table\,\ref{tab_fit}}
    \label{fig_par}
\end{figure}

In the previous section, we successfully model the spectra of obs1--6 with \texttt{MODEL1} and obs7 with \texttt{MODEL2} individually. The best-fit disc inclination angles $i$ for all the epochs are consistent with $i\approx30^{\circ}$ and the disc iron abundances $Z_{\rm Fe}$ are consistent with the solar iron abundance. Therefore, we conduct a multi-epoch spectral analysis by linking the disc iron abundance and the disc inclination angle parameters. \texttt{MODEL1} is used for obs 1-6 and \texttt{MODEL2} is used for obs 7. The best-fit model for each observation is shown in Fig.\,\ref{fig_best_ratio} and the best-fit parameters are shown in Table \ref{tab_fit}. We obtain a very good fit with C-stat/$\nu=10299.69/9299$. A small disc viewing angle of $i=31^{+3}_{-2}$ $^{\circ}$ and a near-solar iron abundance ($Z_{\rm Fe}=1.1\pm0.5Z_{\odot}$) are found. All the observations require a disc density significantly higher than $n_{\rm e}=10^{15}$\,cm$^{-3}$, which was assumed in most of previous reflection-based analysis for other XRB. 

Fig.\,\ref{fig_par} presents the changes of the best-fit parameters with the Eddington ratio. $L_{\rm 0.1-100keV}$ is the 0.1--100~keV band absorption-corrected luminosity calculated using the best-fit model. Only a lower limit of $m_{\rm BH}$ is currently available \citep[e.g. $m_{\rm BH}$>4.9,][]{masetti96}. A black hole mass of 10$M_{\odot}$ is assumed for the calculation of the Eddington luminosity for simplicity. The relative luminosity change between epochs does not change even if this mass assumption is incorrect. Because the same mass value is used for the calculation of the Eddington luminosity for all the epochs.

The first panel of Fig.\,\ref{fig_par} shows the best-fit column density of the X-ray absorption. The upper limit of the Galactic column density along the line of sight is estimated being $3.97\times10^{21}$\,cm$^{-2}$ \citep{willingale13}. Most of the best-fit values obtained in this work are consistent with this value. The column densities obtained for obs 3 and 7 are significantly higher than $4\times10^{21}$\,cm$^{-2}$, possibly indicating a variable line-of-sight neutral absorption. More work concerning possible degeneracy between the column density and other parameters in this analysis can be found in Appendix \ref{appen1} when the thermal emission from the disc is considered in the model for the hard state observations. 

The next four panels of Fig.\,\ref{fig_par} present the evolution of the coronal properties through the outburst. The absolute flux of the coronal emission ($F_{\rm pl}$) increases with the total X-ray flux. The photon index of the coronal emission remains consistent with $\Gamma=1.4$ in obs 1-4, although $F_{\rm pl}$ increases by a factor of 2. As shown in Fig.\,\ref{fig_eeuf}, the coronal emission becomes softer in obs 5 and 6 with $\Gamma=1.5$. The spectra show the softest continuum in obs 7 when the source is in an intermediate state. The high energy cut-off parameter ($E_{\rm cut}$) shows an anti-correlation with the Eddington ratio, except for obs 4 when the source is in the highest flux state. Obs 6, where a softer continuum is found compared to obs 1-4, has the lowest $E_{\rm cut}$, corresponding to a cooler corona. In comparison, obs 7, where the source is in the lowest flux state, has the highest $E_{\rm cut}$, corresponding to a hotter corona. This suggests the temperature of the corona may respond to the accretion rate of the disc. At a higher luminosity/accretion rate, more disc photons are up-scattered in the coronal region, which increases the radiative cooling and lowers the coronal temperature. The fifth panel shows the height of the corona $h$ for each epoch. Only upper limits of $h$ are found. No obvious coorelation between $h$ and the X-ray luminosity is found. Future observations with a higher spectral resolution in the iron band may be able to better constrain the geometry of the corona in BH transients by measuring the emissivity profile of the broad Fe K emission line in more detail.

The variability of the disc reflection is shown in the last three panels of Fig.\,\ref{fig_par}. The absolute flux of the disc reflection component remains consistent when $L_{\rm 0.1-100keV}/L_{\rm Edd}<1\%$ and increases when $L_{\rm 0.1-100keV}/L_{\rm Edd}>1\%$. Due to the decrease in the absolute flux of the coronal emission, obs 7 shows the highest reflection fraction in all the epochs. Only upper limits of the inner radius have been obtained in all the observations. However, a larger upper limit is found at low luminosities, indicating a possible disc truncation (e.g. $r_{\rm in}\approx20r_{\rm g}$) at  $L_{\rm 0.1-100keV}/L_{\rm Edd}=0.5\%$.

\begin{table*}
	\centering
	\caption{The best-fit parameters obtained in the multi-epoch spectral analysis. More details can be found in Section\,\ref{multi}. The flux values of the reflection component ($F_{\rm refl}$) and the coronal emission ($F_{\rm pl}$) are calculated between 1--78~keV in the unit of erg\,cm$^{-2}$\,s$^{-1}$. `l' means that the corresponding parameter is linked in the fit. We assume $d=2.4$\,kpc and $M_{\rm BH}=10M_{\odot}$ for the calculation of the unabsorbed X-ray luminosities and the Eddington luminosity of \src.}
	\label{tab_fit}
	\begin{tabular}{cccccccccc} 
		\hline
		Model & Parameter & Unit & Obs1  & Obs2 & Obs3 & Obs4 & Obs5 & Obs6 & Obs7 \\
		\hline
		\texttt{tbabs} & $N_{\rm H}$ & $10^{22}$\,cm$^{-3}$ & $0.13^{+0.18}_{-0.08}$ & $0.48^{+0.14}_{-0.20}$ & $0.59^{+0.03}_{-0.04}$ & $0.25^{+0.07}_{-0.09}$ & $0.37\pm0.03$ & $0.38^{+0.04}_{-0.03}$ & $0.50\pm0.02$ \\
		\texttt{diskbb} & kT & keV & - & - & - & - & - & - & $0.401^{+0.003}_{-0.002}$\\
		                & norm & - & - & - & - & - & - & - & $6220^{+520}_{-320}$ \\
		\texttt{relconvlp} & h & $r_{\rm g}$ & <11 & <18 & <7 & <12 & <23 & <25 & <20\\
		                   & $R_{\rm in}$ & $r_{\rm g}$ & <14 & <12 & <11 & <8 & <16 & <20 & <30\\
		                   & $i$ & degrees & $31^{+3}_{-2}$ & l & l & l & l & l& l\\
		                   \hline
		\texttt{reflionx}  & $\log(\xi)$ & - & $3.50^{+0.04}_{-0.03}$ & $3.50^{+0.02}_{-0.15}$ & $3.53^{+0.07}_{-0.05}$ & $3.49^{+0.03}_{-0.04}$ & $3.45^{+0.05}_{-0.08}$ & $3.13^{+0.09}_{-0.10}$ & $3.72\pm0.02$ \\
		                   & $Z_{\rm Fe}$ & $Z_{\odot}$ & $1.1\pm0.5$ & l& l& l& l& l& l\\
		                   & $n_{\rm e}$ & $10^{19}$cm$^{-3}$ & $6\pm3$ & $7.1^{+2.4}_{-1.1}$  & $6^{+3}_{-4}$ & $6^{+3}_{-2}$ & $17^{+13}_{-7}$ & $40^{+20}_{-23}$ & $15^{+10}_{-5}$\\
		                   & $\log(F_{\rm refl})$ & erg\,cm$^{-2}$\,s$^{-1}$ & $-8.68^{+0.06}_{-0.02}$ & $-8.67^{+0.07}_{-0.08}$ & $-8.57^{+0.03}_{-0.04}$ & $-8.41^{+0.08}_{-0.03}$ & $-8.51^{+0.03}_{-0.05}$ & $-8.66^{+0.06}_{-0.05}$ & $-8.60^{+0.03}_{-0.04}$\\
		                   \hline
		\texttt{cutoffpl}  & $\Gamma$ & - &  $1.385^{+0.017}_{-0.014}$ & $1.362^{+0.009}_{-0.016}$ & $1.360^{+0.011}_{-0.014}$ & $1.41^{+0.03}_{-0.02}$ & $1.519^{+0.009}_{-0.010}$ & $1.54^{+0.04}_{-0.03}$ & $1.756^{+0.018}_{-0.015}$ \\
		                   & $E_{\rm cut}$ & keV & $160^{+14}_{-12}$ & $143^{+8}_{-10}$ & $140^{+5}_{-6}$ & $162^{+15}_{-7}$ & $134^{+9}_{-8}$ & $99^{+11}_{-5}$ & $200^{+11}_{-7}$\\
		                   & $\log(F_{\rm pl})$ & erg\,cm$^{-2}$\,s$^{-1}$& $-8.071^{+0.009}_{-0.013}$ &  $-8.022^{+0.014}_{-0.016}$ & $-7.97^{+0.011}_{-0.009}$ & $-7.839^{+0.005}_{-0.002}$ & $-8.005\pm0.013$ & $-7.906^{+0.008}_{-0.009}$ & $-8.538^{+0.018}_{-0.029}$ \\
		\hline
		& { $\frac{L_{\rm 0.1-100keV}} {L_{\rm Edd}}$} & \% & 0.81 & 0.90 & 1.00 & 1.19 & 0.99 & 0.97 & 0.50\\
	\end{tabular}
\end{table*}

\section{Discussion} \label{dis}

\begin{figure}
	\includegraphics[width=\columnwidth]{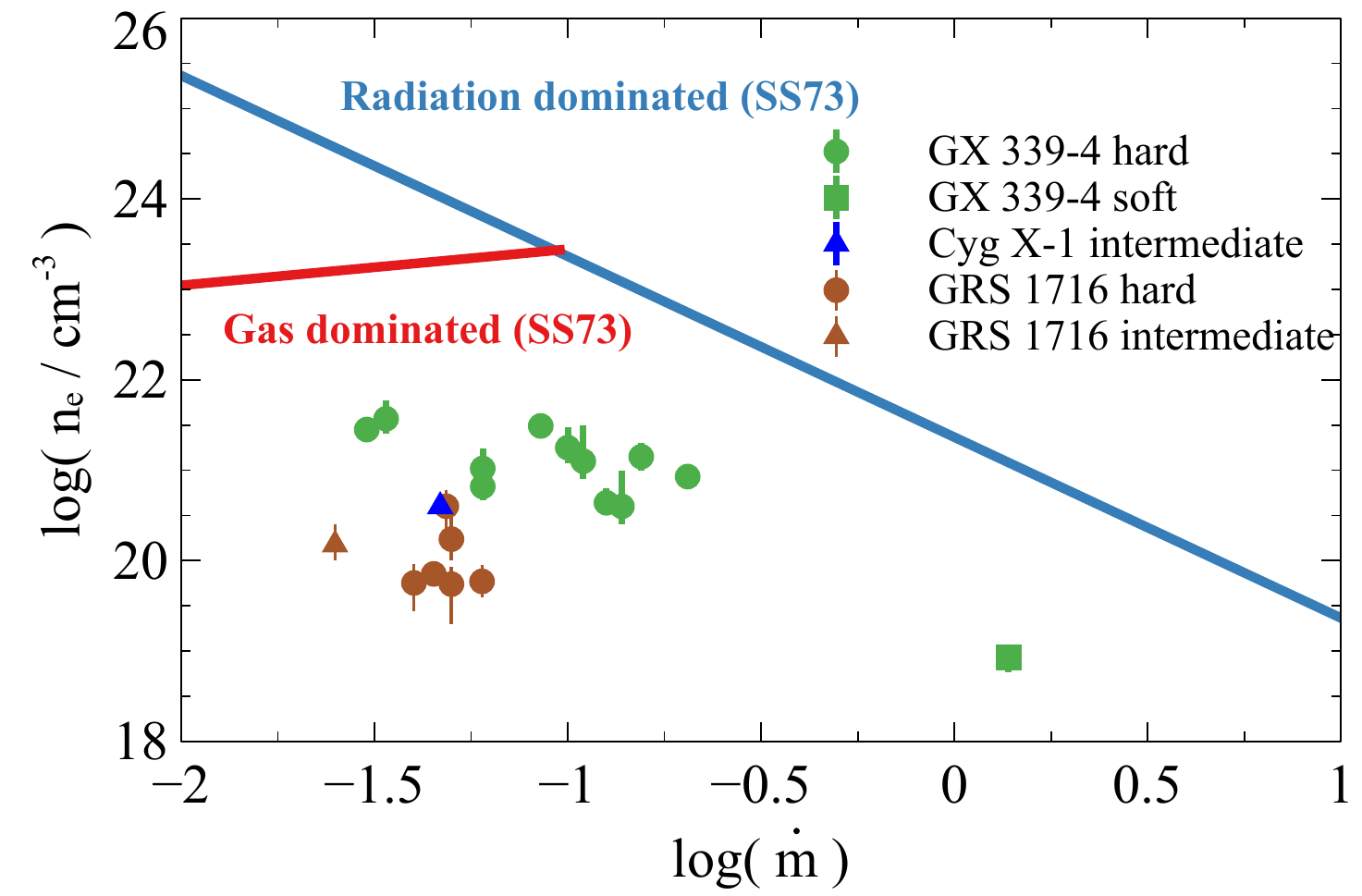}
    \caption{Disc density vs. mass accretion rate $\dot{m}$. Green: GX~339$-$4 \citep{jiang19b}; blue: Cyg~X-1 \citep{tomsick18}; brown: \src. The circles represent hard state observations; the sqaure represents the soft state observation of GX~339$-$4; the triangles represent intermediate state observations. A black hole mass of 10$M_{\odot}$ and an accretion efficiency of 20\% are assumed during the calculation of the mass accretion rates for \src. The blue line is the solution for a radiation pressure-dominated disc at $r=2R_{S}$ calculated by \citet{shakura73}, assuming $r_{\rm in}=1R_{S}$. The red line is the solution for a gas pressure-dominated disc at the same radii of the disc. More details can be found in \citet{jiang19b}.}
    \label{fig_ne}
\end{figure}

We have undertaken an analysis of the broadband (1--78 keV) X-ray spectra of the BH XRB \src\ taken by \nustar\ and \swift\ during its recent `failed' outburst. Following our recent work on GX~339$-$4 \citep{jiang19b}, we focus on modeling the data with reflection models that allow the density of the disc to be fit as a free parameter \citep[see also][]{tomsick18}.

We first summarize the spectral fitting results: the first four observations were taken at the rising phase of the outburst. The spectra show a consistent spectral shape but an increasing X-ray luminosity (from 0.8\% to 1.2\% of $L_{\rm Edd}$). The inner radius of the disc is found to be $<8r_{\rm ISCO}$ in obs 4 when the source reaches the highest X-ray luminosity, indicating a disc extending to the ISCO or a slightly truncated disc. The spectra of obs 5 and 6 become softer than obs 1-4 although the source remains at a similar X-ray luminosity as the previous observations. During this transition, obs 6 shows the lowest high energy cut-off in the coronal emission, indicating a cool coronal temperature (e.g. T=30--50~keV\footnote{The value of $E_{\rm cut}$ is approximately 2--3 times of $kT_{\rm e}$, depending on the optical depth of the corona.}). At the same time, the density in the surface of the disc increases by a factor of 5 compared to the density in obs 1-4. The last observation, obs 7, was taken 7 months after the beginning of the outburst. The spectra of obs 7 show a combination of a disc thermal emission ($kT=0.4$keV), a softer coronal emission ($\Gamma=1.76$), and a strong disc reflection component, indicating the source is in a very low flux intermediate state. \src\ shows the lowest X-ray luminosity in obs 7 compared to obs 1-6, but has a similar absolute flux from the disc reflection component to these higher flux observations, indicating a possible change in the geometry of the corona. However only upper limits of the coronal height are obtained. 

Second, we discuss the disc density measurements during the outburst of \src. The disc densities measured in this work are shown in brown in Fig.\,\ref{fig_ne}. The green circles represent the observations of GX~339$-$4 in the hard state, and the green circle represents the high flux-soft state observations of the same source in 2015 \citep{jiang19b}. The blue triangle shows the disc density of Cyg~X-1 in the intermediate state \citep{tomsick18}. The mass accretion rate $\dot{m}$ is estimated using $\dot{m}=L_{\rm Bol}/\epsilon L_{\rm Edd}\approx L_{\rm 0.1-100keV}/\epsilon L_{\rm Edd}$, where $\epsilon$ is the accretion efficiency. A black hole mass of $10M_{\odot}$ and an accretion efficiency of $\epsilon=20\%$ are assumed in our calculation. Assuming the distance measurements of GX~339-4 and \src\ are accurate and the BH masses used for these calculations are correct, the hard states of two sources show a similar Eddington ratio. However the disc density in \src\ is 10 times lower than the density in GX~339-4 \citep{jiang19b}. The difference of the disc density could be due to the different BH mass, the accretion efficiency, and the true intrinsic bolometric luminosity. The intermediate state of \src\ shows a similar disc density as the intermediate state of Cyg~X-1 but a much lower luminosity in the X-ray band. Nevertheless, we conclude that the disc density required for the broad band spectral analysis of \src\ is significantly larger than $n_{\rm e}=10^{15}$\,cm$^{-3}$, which was assumed in previous reflection-based spectral modelling.

The blue solid line in Fig.\,\ref{fig_ne} shows the disc density at $r=2R_{\rm S}$ of a radiation pressure-dominated disc varying with the parameter $\dot{m}$. $r_{\rm in}=1R_{\rm S}$ is assumed for the disc inner radius. We refer interested readers to \citet{svensson94} for more details and \citet{jiang19b} for the complete equations. In this thin disc model, the density of the disc has an anti-correlation with the BH mass accretion rate ($n_{\rm e} \propto \dot{m}^{-2}$). However, a significantly lower disc density has been found in various BH transients \citep[e.g.][]{tomsick18,jiang19b} compared to the radiation pressure-dominated disc model. The reasons of the offset might be 1) the uncertainties of the BH mass and distance measurements. This may also be able to explain a lower disc density in \src\ than in GX~339-4 due to the underestimation of $\dot{m}$. We note, however, that the distance and the BH mass of Cyg~X-1 is well constrained but the density given by reflection spectroscopy is still lower than the classical disc model; 2) the low accretion rate ($\dot{m}\lessapprox1\%\dot{m}_{\rm Edd}$) in the hard state suggests that the disc might be dominated by gas pressure in the inner region instead of radiation pressure \citep{shakura73,svensson94}. The vertical structure of a gas pressure-dominated disc can be estimated by assuming an isothermal model \citep[e.g.][]{frank02}. For a stellar-mass BH, such as GX~339$-$4 and \src, the disc density within one Thomson optical depth in the surface of the disc is approximately 100 times lower than the density in the mid-plane \citep{jiang19c}. However, the disc density in the standard disc density model shown in Fig.\,\ref{fig_ne} is calculated assuming a uniform density across the height of a thin disc \citet{shakura73,svensson94}. This assumption has been proven to be invalid in later studies of discs \citep[e.g.][]{ionson84,turner04}.

\section*{Acknowledgements}

J.J. acknowledges support by the Cambridge Trust and the Chinese Scholarship Council Joint Scholarship Programme (201604100032), the Tsinghua Shuimu Scholar Programme and the TAO Fellowship. D.J.W. acknowledges support from an STFC Ernest Rutherford Fellowship. A.C.F. acknowledges support by the ERC Advanced Grant 340442. M.L.P. and F.F. are supported by European Space Agency (ESA) Research Fellowships. This work was based on the \texttt{reflionx} code written by Randy R. Ross. We acknowledge his big contribution to this project. This work made use of data from the NuSTAR mission, a project led by the California Institute of Technology, managed by the Jet Propulsion Laboratory, and funded by NASA. This research has made use of the NuSTAR Data Analysis Software (NuSTARDAS) jointly developed by the ASI Science Data Center and the California Institute of Technology. This work made use of data supplied by the UK Swift Science Data Centre at the University of Leicester. 




\bibliographystyle{mnras}
\bibliography{grs.bib} 




\appendix

\section{Constraining the disc thermal component} \label{appen1}

\begin{figure*}
    \centering
    \includegraphics[width=18cm]{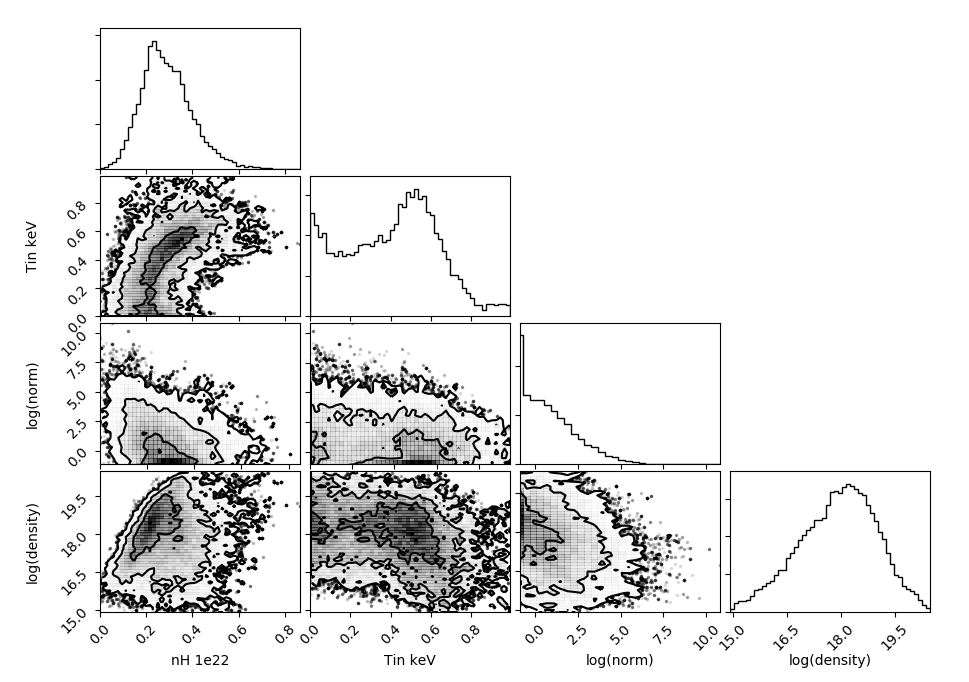}
    \caption{Output distributions for the MCMC analysis of the best-fit model of the broad band obs 4 spectra of \src. Both \nustar\ and \swift\ spectra are considered. The model includes a relativistic reflection model with a variable disc density parameter and a disc thermal component. Only some of the parameters that affect the soft spectral modelling are shown here for clarity. The `norm' label represents the normalization of the \texttt{diskbb} model. Contours correspond to 1, 2 and 3$\sigma$.}
    \label{pic_contour}
\end{figure*}

\begin{table*}
    \centering
    \begin{tabular}{ccccc}
    \hline\hline
    Model & Parameter & Unit & \texttt{MODEL2} & \texttt{MODEL1} \\
    \hline
    \texttt{tbabs} & $N_{\rm H}$ & $10^{21}$\,cm$^{-2}$ & $2.5\pm1.5$ & $2.5\pm0.6$\\
    \texttt{diskbb} & kT & keV & <0.85 & -\\
                    & norm & - & <$5\times10^{5}$ & - \\
    \texttt{relconvlp} & h & $r_{\rm g}$ & <18 & <13\\
                       & $R_{\rm in}$ & $r_{\rm g}$ & <6 & <7 \\
                       & $i$ & degrees & $30^{+12}_{-8}$ & $32^{+7}_{-5}$\\
    \texttt{reflionx} & $\log(\xi)$ & - & $3.50^{+0.20}_{-0.12}$ & $3.41\pm0.05$\\
                      & $Z_{\rm Fe}$ & $Z_{\odot}$ & $1.4^{+2.1}_{-0.6}$ & $1.2^{+0.7}_{-0.3}$\\
                      & $\log(n_{\rm e})$ & cm$^{-3}$ & $19.2^{+0.4}_{-3.5}$ & $19.77\pm0.05$\\
                      & $\log(F_{\rm refl})$ & erg cm$^{-2}$ s$^{-1}$ & $-8.38^{+0.02}_{-0.10}$ & $-8.42\pm0.03$\\
    \texttt{cutoffpl} & $\Gamma$ & - & $1.41\pm0.02$ & $1.41^{+0.02}_{-0.03}$\\
                      & $E_{\rm cut}$ & keV & $152^{+8}_{-13}$ & $158^{+12}_{-9}$\\
                      & $\log(F_{\rm pl})$ & erg cm$^{-2}$ s$^{-1}$ & $-7.85\pm0.02$ & $-7.838^{+0.006}_{-0.005}$\\
                      & C-stat/$\nu$ & -& 1387.76/1154 & 1389.12/1156 \\
    \hline
    \end{tabular}
    \caption{The best-fit model parameters obtained by fitting the spectra of obs 4 with \texttt{MODEL1} and \texttt{MODEL2}. \texttt{MODEL2} includes an additional \texttt{diskbb} component that accounts for the disc thermal spectrum.}
    \label{tab_obs4_fit}
\end{table*}

Previous spectral modelling of BH transients in the hard state commonly requires an additional weak disc-blackbody component when assuming a fixed disc density parameter ($n_{\rm e}=10^{15}$\,cm$^{-3}$) for the reflection model. The origin of such a disc-blackbody component remains unclear. For example, the hard state of GX~339$-$4 shows $L_{\rm X} \lessapprox 1\% L_{\rm Edd}$ and the spectral modelling suggests the presence of a disc-blackbody component with a temperature of $kT > 0.8$\,keV \citep[e.g.][]{jingyi18}. Such a high temperature is comparable to the one in the soft state of the same source, where the X-ray band is dominated by the disc thermal component and has a luminosity of $L_{\rm X} \approx 20\% L_{\rm Edd}$ \citep[e.g.][]{parker16,jiang19b}. 

In this work, the blackbody-shaped emission shown in the hard state of \src\ is instead interpreted as part of a high density disc reflection component. A disc density of $10^{19}-10^{20}$\,cm$^{-3}$ is found using high density disc reflection spectroscopy. See Section \ref{multi} for more details. Similarly, \citet{jiang19b} analysed the \nustar\ and \swift\ spectra of GX~339$-$4 in the hard state and found that the disc thermal component is statistically degenerate with the column density when both disc thermal component and high density disc reflection are considered. In this section, we apply the same method as in \citet{jiang19b} to the hard state spectra of \src. Both relativistic disc reflection and disc thermal components are considered (\texttt{MODEL2}). We do not consider obs 1--3 as they do not have soft X-ray coverage. Only obs 4 is shown here as an example. Similar conclusions can be found for the rest of the hard-state observations. Note that the \swift\ and \nustar\ spectra of obs 4 cover 1--78~keV. The \swift\ spectrum below 1~keV is ignored due to calibration issue. 

The best-fit parameters given by \texttt{MODEL2} are shown in Table\,\ref{tab_obs4_fit}, in comparison with the parameters given by \texttt{MODEL1} where only reflection is considered. \texttt{MODEL2} gives a slightly worse fit compared to \texttt{MODEL1} with $\Delta $C-stat=5 and two more free parameters. In conclusion, both two models offer similar good fits to the spectra. However, only upper limits of the disc density and the disc thermal component are obtained when an extra \texttt{diskbb} model is added. The column density shows a larger measurement uncertainty calculated by using the ERROR command in XSPEC when \texttt{MODEL2} is used, although they are consistent with the results given by \texttt{MODEL1}. Different uncertainty ranges given by \texttt{MODEL2} likely indicate parameter degeneracy when both reflection and disc thermal component are considered. Therefore, we checked the constraints of all the parameters by using the MCMC algorithm. The XSPEC/EMCEE code by Jeremy Sanders based on the python implementation  \citep{foreman12} and the MCMC ensemble sampler \citep{goodman10} was used. We use 100 walkers with a length of 400000, burning the first 5000. A convergence test has been conducted and the Gelman-Rubin scale-reduction factor $R<1.3$ for every parameter. Fig.\,\ref{pic_contour} shows the output distributions of some of the parameters that affect the soft X-ray spectral modelling. 

From our MCMC analysis, we find that: 
\begin{itemize}

    \item The column density $N_{\rm H}$ of the line-of-sight absorption towards \src\ is well constrained between $(0.5-6)\times10^{21}$\,cm$^{-2}$ (3$\sigma$). This result is consistent with the upper limit of the Galactic absorption towards \src\ \citep[e.g. $N_{\rm H}=4.7\times10^{21}$\,cm$^{-2}$,][]{willingale13}. However, $N_{\rm H}$ shows degeneracy with the temperature of the disc blackbody component $kT$ and the disc density parameter $n_{\rm e}$. When $kT$ and $n_{\rm e}$ are higher, the continuum model produces a stronger emission in the soft X-ray band and thus a higher $N_{\rm H}$ is required to fit the spectrum. 
    
    \item When $kT$ is lower, a higher value of $n_{\rm e}$ is required to model the blackbody emission in the soft X-ray band. For example, our MCMC analysis shows that 1) $n_{\rm e}$ is in the range of $10^{16}-10^{20}$\,cm$^{-3}$ ($3\sigma$) when $kT<0.4$\,keV; 2) in comparison, $n_{\rm e}$ is in the range of $10^{15}-10^{19.5}$\,cm$^{-3}$ ($3\sigma$) when $kT\approx0.5$\,keV. The former fit is slightly preferred than the latter fit according to our MCMC analysis, although they have similar likelihood within the 2$\sigma$ uncertainty range. 
    
    \item Only an upper limit of $kT$ is obtained due to the lack of data below 1\,keV. Future observations, such as from \textit{NICER}, will be able to help constraining the soft X-ray spectral shape in BH XRBs with more details.
\end{itemize}

To sum up, we can only obtain an upper limit of the contribution of the disc thermal component in the hard state of \src\ due to the modest column density along the line of sight and the lack of data below 1\,keV. The disc thermal component is also degenerate with the density parameter in the high density disc reflection model used in this work. We note that when $n_{\rm e}=10^{15}$\,cm$^{-3}$ is used, a disc blackbody component with $kT\approx0.5$\,keV is required to model the soft X-ray emission. This result is consistent with the analyses in \citet{bassi19}. 



\bsp	
\label{lastpage}
\end{document}